\documentclass[twoside,leqno,twocolumn]{article}

\usepackage[letterpaper]{geometry}

\usepackage{ltexpprt}
\usepackage{hyperref}

\usepackage{graphicx}
\usepackage{subcaption}
\usepackage{float} 
\usepackage{afterpage}
\usepackage{bm}
\usepackage{colortbl}
\usepackage{titlesec}
\usepackage{multirow}
\usepackage{tablefootnote}
\usepackage{amsmath} 
\usepackage[table,xcdraw]{xcolor}
\usepackage{booktabs}

\usepackage{amsthm}
\theoremstyle{definition}
\newtheorem{definition}{Definition}
\usepackage{mathrsfs}

\definecolor{lightgray}{cmyk}{0, 0, 0, 0.173}

\newenvironment{keywords}{
    \noindent\textbf{Keywords:}
}
{}

\begin{document}

%


\title{\Large A Look Into News Avoidance Through \textit{AWRS}: An Avoidance-Aware Recommender System
}

\author{Igor L.R. Azevedo\footnotemark[1] \and Toyotaro Suzumura \thanks{The University of Tokyo, Tokyo, Japan,\\ Email: \{lima-rocha-azevedo-igor, suzumura\}@g.ecc.u-tokyo.ac.jp} \and Yuichiro Yasui \thanks{Nikkei Inc., Tokyo, Japan\\ Email: yuichiro.yasui@nex.nikkei.com}}

\date{}

\maketitle


\fancyfoot[R]{\scriptsize{Copyright \textcopyright\ 2025 by SIAM\\
Unauthorized reproduction of this article is prohibited}}



\fancyfoot[R]{\scriptsize{Copyright \textcopyright\ 2025 by SIAM\\
Unauthorized reproduction of this article is prohibited}}

\pagenumbering{arabic}
\setcounter{page}{1}

\begin{abstract} \small\baselineskip=9pt 
In recent years, journalists have expressed concerns about the increasing trend of news article \textit{avoidance}, especially within specific domains. This issue has been exacerbated by the rise of recommender systems. Our research indicates that recommender systems should consider avoidance as a fundamental factor. We argue that news articles can be characterized by three principal elements: \textit{exposure}, \textit{relevance}, and \textit{avoidance}, all of which are closely interconnected. To address these challenges, we introduce \textit{AWRS}, an Avoidance-Aware Recommender System. This framework incorporates avoidance awareness when recommending news, based on the premise that \textit{news article avoidance conveys significant information about user preferences}. Evaluation results on three news datasets in different languages (English, Norwegian, and Japanese) demonstrate that our method outperforms existing approaches.
\end{abstract}


\section{Introduction.}

Recommender systems are extensively used to present users with options that align closely with their interests. In the news domain, these systems face unique challenges not typically encountered in other areas, such as the importance of \textit{timeliness}, \textit{novelty}, and \textit{relevance}. 


Traditional methods based on user click history have evolved into advanced techniques that better capture user preferences and behaviors. The NRMS model \cite{wu-etal-2019-neural-news}, for example, uses multi-head self-attention to learn news representations from titles and to capture user preferences from their browsing history. The NAML model \cite{wu2019neural} utilizes convolutional neural networks (CNNs) and enriches user and news representations by leveraging various types of news information, such as titles, bodies, and categories. The LSTUR model \cite{an-etal-2019-neural} focuses on both short- and long-term user representations and selects important words through an attention network. Recent models have further advanced performance through new approaches. The GLORY model \cite{Yang_2023} integrates global information with local user interactions for better personalization. LANCER \cite{Bae_Ahn_Lee_Kim_2023} uses the concept of news lifetime to enhance negative sampling by considering the finite influence period of news. The PP-Rec model \cite{qi2021pprec} tackles cold-start and popularity bias by combining personalized and popularity-based ranking scores. CAUM \cite{qi2022news} leverages candidate-aware self-attention networks to capture global user interests and incorporates candidate news into local context modeling.

A growing trend in today's digital landscape is \textit{news avoidance}, where users deliberately reject or neglect certain news, influenced by a preference for alternative media \cite{what_do_news_readers, taking_break_from_news_articles, no_news_is_not_good}. This behavior, often temporary or selective, highlights the need for recommender systems to integrate avoidance patterns. According to \cite{no_news_is_not_good}, mental health concerns are identified as the primary factor driving the increase in news avoidance, with growing distrust in mainstream media also being a significant contributor. Although skepticism toward journalism is not a new issue, it was amplified during the pandemic by the influence of anti-vaccination advocates and conspiracy theorists. Despite advancements in news modeling for recommendation systems, the integration of \textit{news avoidance} behaviors remains largely unexplored.

In summary, our contributions are as follows: \textbf{(1)} Our model presents a novel concept, and to the best of our knowledge, we are the first to explore the perspective of \textit{avoidance} in recommender systems. \textbf{(2)} We introduce the \textit{AWRS} framework (Avoidance-Aware Recommender System), which incorporates avoidance awareness, including time and relevance modules, to improve performance in user matching recommendations. \textbf{(3)} Extensive experiments on three diverse real-world datasets show that \textit{AWRS} consistently delivers superior performance across a wide range of metrics.

\section{Related Work.}\label{subsec:rec_aspects}

Neural networks models have emerged as the leading approach for personalized news recommendations, leveraging news and user encoders to compute recommendation scores by comparing candidate news embeddings with user embeddings \cite{10.1145/3530257, wu-etal-2019-neural-news, Wu_2019, iana2024train, 10.1080/21670811.2021.2021804, Gharahighehi_2021, Moreira_2019, qi2022news}. 

Recent advancements in news recommender systems have primarily focused on four key aspects: (1) \textbf{\textit{Sentiment}} – SentiRec \cite{wu-etal-2020-sentirec}, which recommends news with diverse sentiment profiles; (2) \textbf{\textit{Popularity}} – PP-Rec \cite{qi2021pprec}, which leverages news popularity to address cold-start issues and enhance diversity; (3) \textbf{\textit{Global Representations}} – GLORY \cite{Yang_2023}, which enriches news representations by utilizing a global news graph and gated graph neural networks; and (4) \textbf{\textit{Recency}} – LANCER \cite{Bae_Ahn_Lee_Kim_2023}, which incorporates the concept of news lifetime to improve recommendation accuracy. 

In this work, we introduce a novel perspective on news recommendation by integrating \textbf{\textit{avoidance awareness}}, aiming to enhance the understanding of user engagement patterns and preferences.


\section{Principal Elements.}

\begin{figure}[htp]
  \centering
  \includegraphics[width=0.9\linewidth]{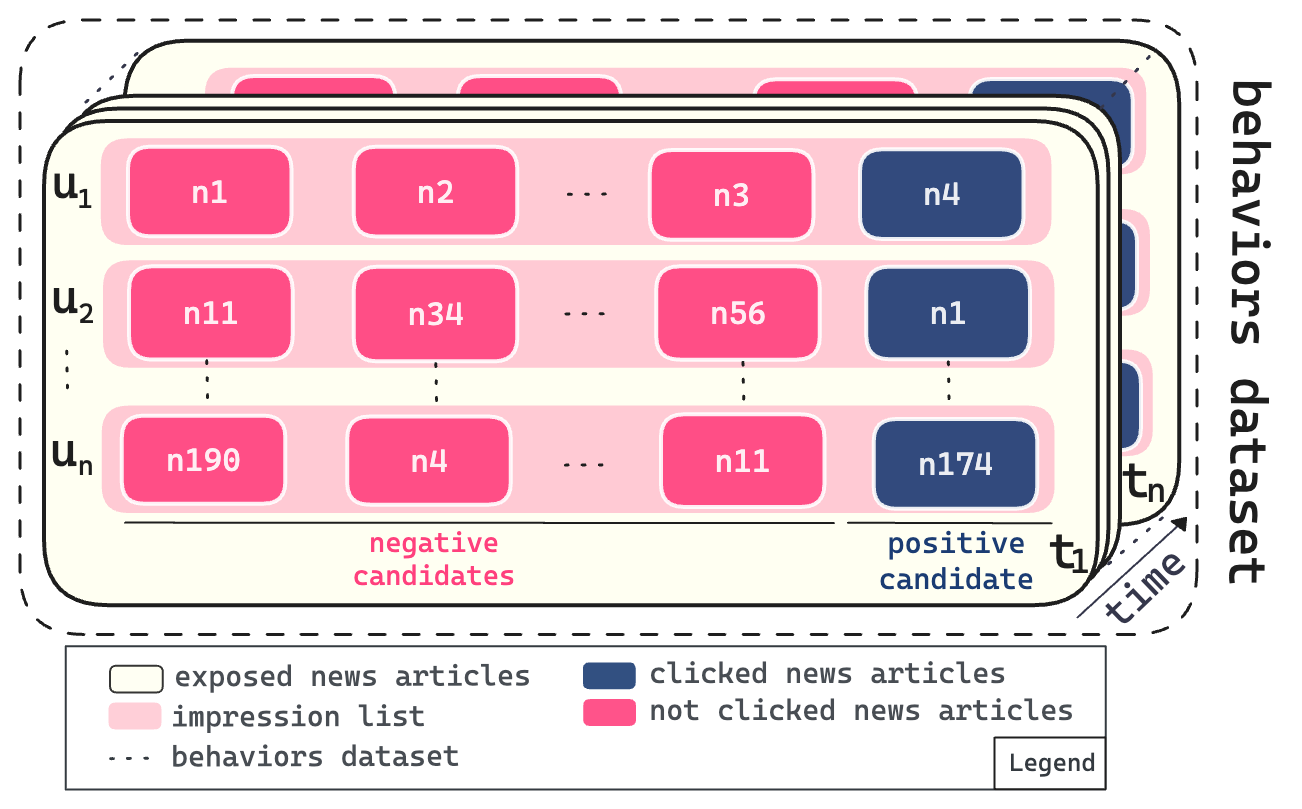}
  \caption{\textit{Avoidance} and \textit{exposure} explanation diagram}
  \label{fig:explaining_avoidance}
\end{figure}

We identified three principal elements for characterizing a news article: \textit{exposure}, \textit{avoidance}, and \textit{relevance}. Before defining these concepts, please refer to Figure \ref{fig:explaining_avoidance}, which is needed to illustrate some terms. In the figure, the black dashed line represents the behaviors dataset, listing news articles (impressions) at each time point \( t \). For simplicity, assume that each clicked news article (positive candidate) has a fixed number of non-clicked articles (negative candidates). The section labeled ``\textit{exposed news articles}" includes all the news articles viewed by users at time \( t_1 \). Consider a news platform where, at any given time \( t \), a set of news articles \(\mathcal{N}_t = \{n_1, n_2, \ldots, n_k\}\) and a group of users \(\mathcal{U}_t = \{u_1, u_2, \ldots, u_n\}\) interact. When a user \( u_1 \) views an article \( n_1 \), they are ``exposed" to it and can choose to click on it or not. This section contains all the articles \(\mathcal{N}\) viewed by all users \(\mathcal{U}\) at a specific time \( t \).

The “\textit{impression list}” represents all news articles shown to a user \( u \), including both those that were clicked and those that were not. An impression log records the articles displayed to a user at a specific time and their click behaviors \cite{wu-etal-2020-mind}. The set of impressions at time \( t \) is referred to as \textit{exposure}. Exposure depends on the publication date (time) of news articles; only articles published before the exposure time can be displayed to users. Therefore, at a specific time \( t \), some news articles may be exposed while others are not, simply because they have not yet been published. This concept is illustrated by the arrow at the bottom of the figure, which shows how the impression lists change over time.  


\subsection{Exposure.}

\begin{definition}[\textbf{Number of Exposures - \( n_E(n, t) \)}]\label{def:exposure}
    The total number of times a specific news article \( n \) is presented to users on the platform up to a defined time \( t \). This metric quantifies all instances where the article \( n \) was displayed to active users, thus providing opportunities for being clicked.
\end{definition}

\begin{definition}[\textbf{Number of Impressions - \( n_I(t) \)}]\label{def:impression}
    An \textit{Impression} is defined as an individual data record in a news dataset in a given time $t$. 
    
\end{definition}

\begin{definition}[\textbf{Exposure Per Impression - \(EPI(n, t)\)}]\label{def:epi}
    The Exposure Per Impression (\(EPI\)) for a news article \(n\) at time \(t\) is calculated as follows:
    
    \vspace{-1.5em}
    \begin{equation}
        EPI(n, t) = \frac{n_E(n, t)}{n_I(t)}
    \end{equation}
    
    This ratio provides a relative measure of how extensively a news article \(n\) has been exposed in relation to the total number of impressions recorded.
\end{definition}

\noindent \textit{\underline{Example}}: At time \(t_1\), the news article \(n_{174}\) appears in 50 out of 100 impression lists, giving \(n_E(n_{174}, t_1) = 50\) and \(n_I(t_1) = 100\). Thus, its exposure per impression is \(EPI(n_{174}, t_1) = \frac{50}{100} = 0.5\).

\subsection{Avoidance.} Avoidance represents a reader's decision to disengage from specific news content due to factors such as disinterest, discomfort, or distrust. Analyzing avoidance patterns is key to understanding user behavior and optimizing content recommendations.

\begin{definition}[\textbf{Avoidance - \( Av(n, t) \)}]\label{def:avoidance}
    Avoidance for a news article \( n \) over a timeframe \( t \) quantifies the proportion of potential exposures that did not result in interaction. This is mathematically defined as follows:

    \vspace{-1.5em}
    \begin{equation}
        Av(n, t) = 1 - \frac{n_{clk}(n, t)}{n_E(n, t)}
    \end{equation}

    \noindent here, \( n_{clk}(n, t) \) represents the number of clicks received by the article \( n \) up to the specified time \( t \), and \( n_E(n, t) \) denotes the total number of exposures of the article \( n \) within the same timeframe. The value of \( Av(n, t) \) ranges from 0 to 1, where 1 indicates \textit{total avoidance} (no clicks relative to exposures) and 0 signifies \textit{complete engagement} (every exposure resulted in a click).
\end{definition}

\noindent \textit{\underline{Example}}: The avoidance of news article \(n_{174}\) at time \(t_1\), with 20 clicks from 50 exposures, is $Av(n_{174}, t_1) = 1 - \frac{n_{clk}(n_{174}, t_1)}{n_E(n_{174}, t_1)} = 1 - \frac{20}{50} = 0.6$.

\subsubsection{Consequences of Avoiding News Articles.}

Consider a large set of users exposed to a news article $n_{174}$ who chose not to click on it, either deliberately or unintentionally, indicating a lack of interest. Now, suppose a specific user, $u_5$, does click on it. Since most users avoided the article, $u_5$'s engagement reveals deeper insights into their preferences compared to interactions with widely popular articles. Thus, engaging with avoided articles can better reflect a user's unique interests, while avoidance serves as a measure of disinterest or resistance to a topic.

\subsubsection{Visualization of Avoidance.} 
\begin{figure}[h]
  \centering
  \includegraphics[width=0.85\linewidth]{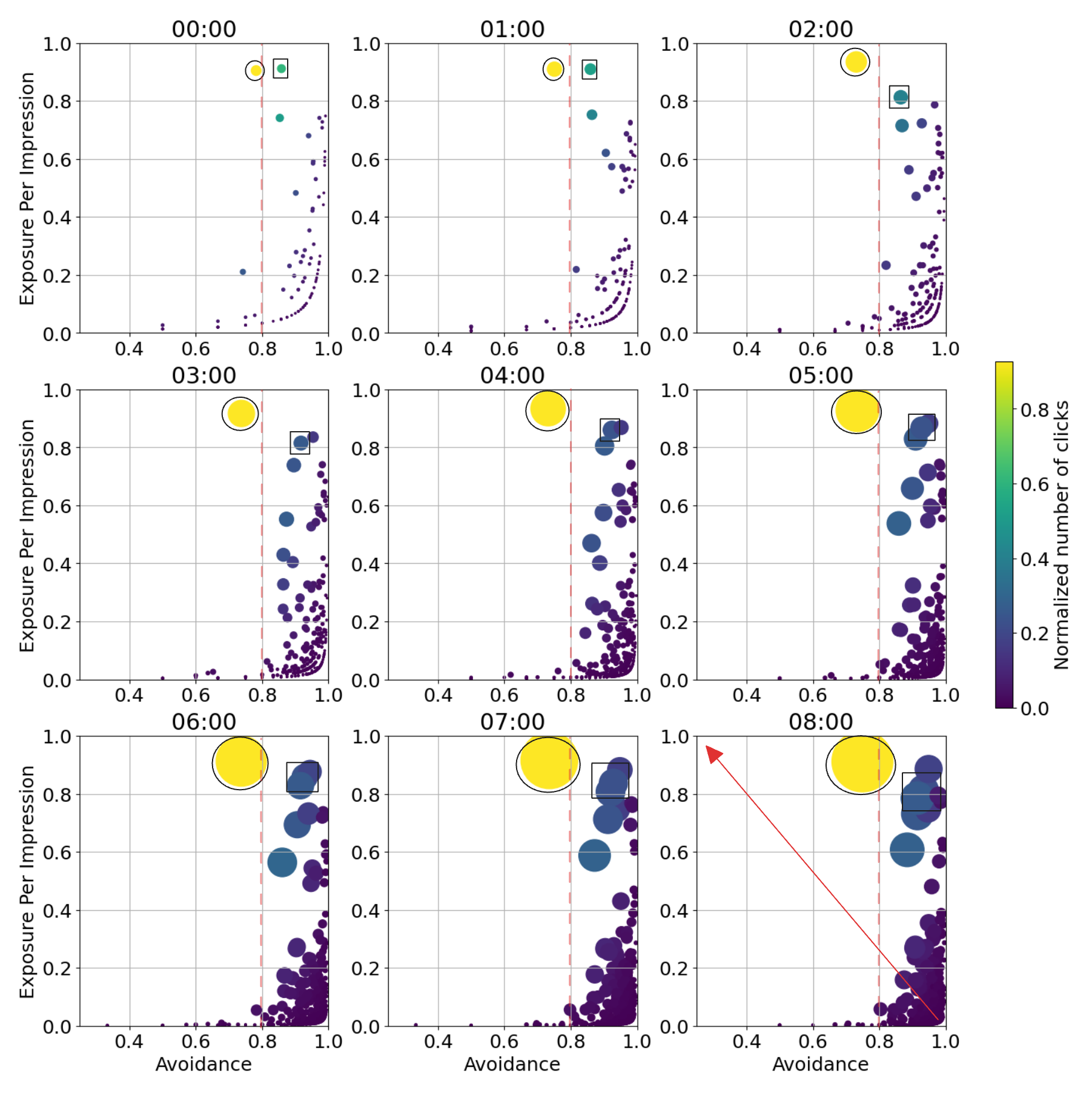}
  \caption{\(Av(n, t) \text{ vs. } EPI(n, t)\) for \textit{MIND-small} \cite{wu-etal-2020-mind} at 2019-11-09 from 00:00 AM to 08:00 AM.}
  \label{fig:graph_usr_eng}
\end{figure}

We analyzed the correlation between news article \textit{avoidance}, \textit{exposure}, and the number of clicks per article. Figure \ref{fig:graph_usr_eng} shows \( Av(n, t) \) versus \( EPI(n, t) \) for the \textit{MIND-small} \cite{wu-etal-2020-mind} dataset on November 9, 2019, from 00:00 AM to 08:00 AM. The color scale represents the normalized click count, and the radius of each article reflects its number of clicks, scaled by a factor of three for better visualization.

Note how the overall number of clicks increases over time, as expected, since more users log into the platform. Additionally, we have highlighted two specific articles: one with a black square and the other with a black circle. Notably, the article marked with the black circle consistently received more clicks over time compared to all other articles. This article maintains a high \( EPI \) and low \( Av \), with \( Av(n, t) \) remaining below 0.8 (red dashed line) at all times, maintaining the highest normalized number of clicks.

In contrast, the article marked with the black square starts with a green color, but as time progresses, its avoidance increases while its exposure per impression decreases. As a result, its relative number of clicks gradually declines, with the color shifting from green to blue. This suggests that articles exhibiting an increase in \( EPI \) and a decrease in \( Av \) (as indicated by the red arrow in the bottom-right chart) tend to receive the highest number of clicks.

This analysis demonstrates how the dynamics of \textit{avoidance} and \textit{exposure} influence click behavior. For a detailed hour-by-hour visualization of avoidance changes, please refer to Appendix \ref{sec:appendix_hourly}.

\begin{figure*}[htb]
    \centering
    \begin{subfigure}[b]{0.32\linewidth}
        \centering
        \includegraphics[width=\linewidth, height=5.0cm]{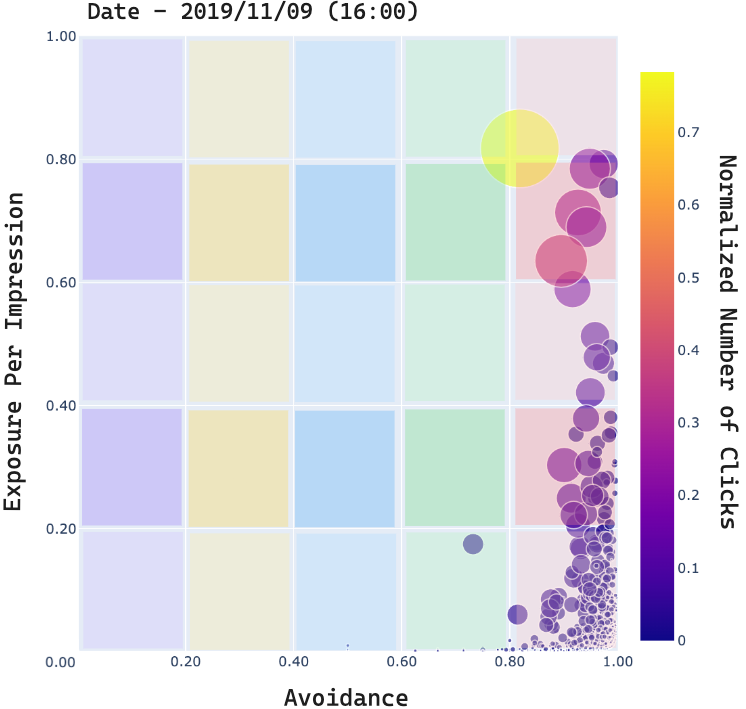}
        \caption{\textit{MIND-small} - English}
        \label{subfig:grid_mindsmall}
    \end{subfigure}
    \hfill
    \begin{subfigure}[b]{0.32\linewidth}
        \centering
        \includegraphics[width=\linewidth, height=5.0cm]{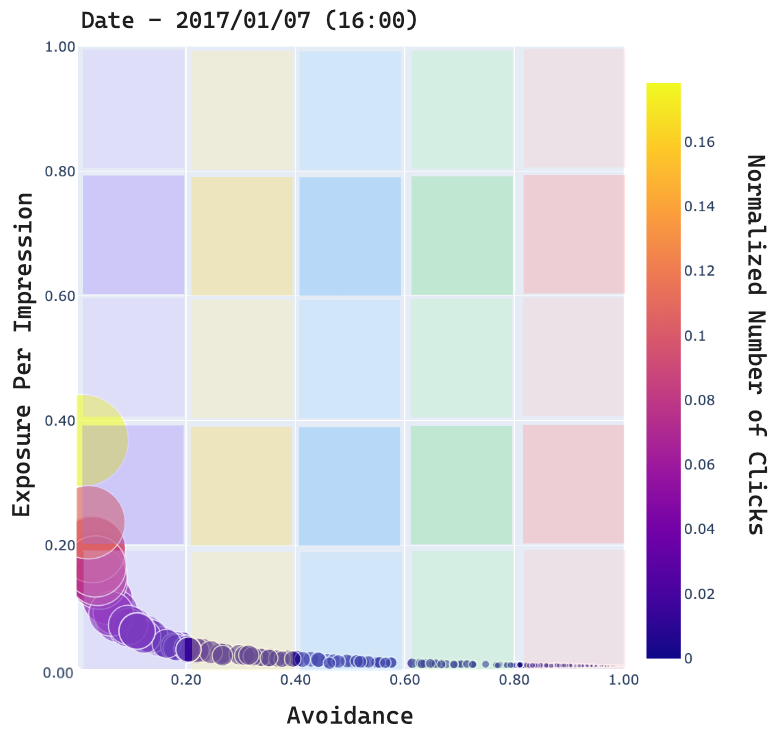}
        \caption{\textit{Adressa one-week} - Norwegian}
        \label{subfig:grid_adressa}
    \end{subfigure}
    \hfill
    \begin{subfigure}[b]{0.32\linewidth}
        \centering
        \includegraphics[width=\linewidth, height=5.0cm]{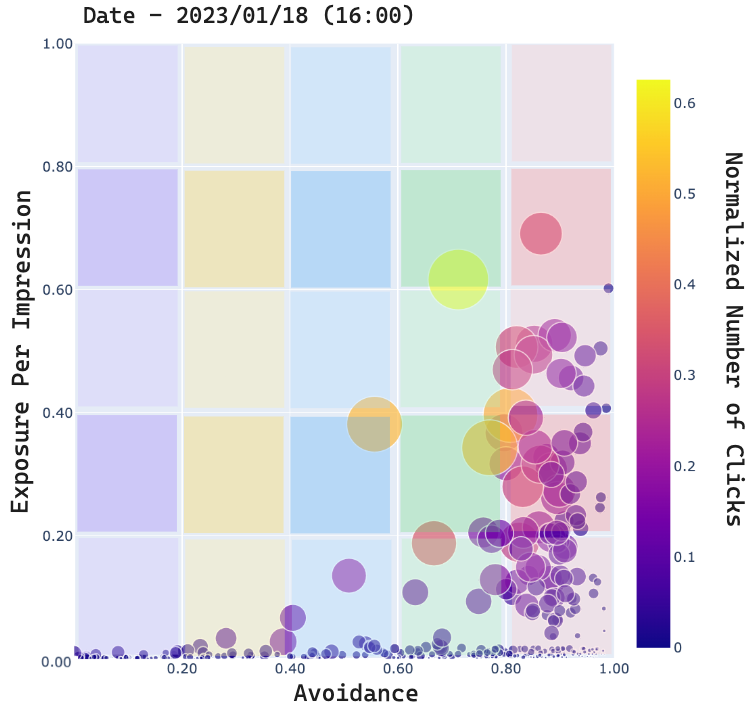}
        \caption{\textit{Nikkei one-week} - Japanese}
        \label{subfig:grid_JPNewsSite}
    \end{subfigure}

    \caption{The graph of \( Av(n, t) \) versus \( EPI(n, t) \) is generated for the \textit{MIND-small} \cite{wu-etal-2020-mind} (a), \textit{Adressa one-week} \cite{10.1145/3106426.3109436} (b), and \textit{Nikkei one-week} (c) datasets, using \( D = 5 \), which results in \textbf{25} distinct regions.}
    \label{fig:grid}
\end{figure*}

\begin{figure}[htb]
  \centering
  \includegraphics[width=\linewidth]{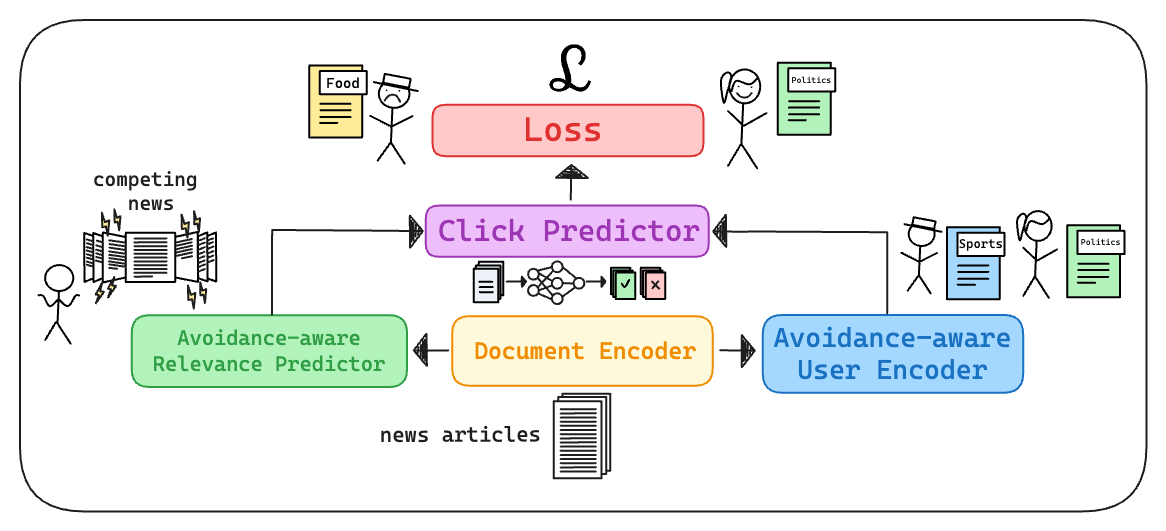}
  \caption{Overview of \textit{AWRS}.}
  \label{fig:awrs_schema}
\end{figure}

\subsection{Relevance.} \label{sec:relevance}

Indicates how well a news article matches reader interests, requiring specific data like (1) time since publication, (2) text embedding, (3) exposure per impression \( EPI(n, t) \), (4) avoidance \( Av(n, t) \), and (5) clicks \( clk(n, t) \). The model weighs these factors to assess relevance, aligning news with user interests. See Section \ref{sec:aw_rel} for details on relevance calculation.

\section{User Engagement Embedding.}\label{sec:usr_eng}

To measure the impact of \textbf{avoidance} and \textbf{exposure} on user interaction with a news article, we calculate a metric called \textit{user engagement} (\(\bm{ue}\))—how users interact with a specific news article based on its \textit{avoidance} and \textit{exposure per impression}. We divide our graph of \(Av(n, t) \text{ vs. } EPI(n, t)\) into $D \times D$ distinct regions. Depending on the values of \(Av(n, t)\) and \(EPI(n, t)\), a news article falls into different regions. 

Formally, the user engagement indices are computed as:

\vspace{-1.5em}
\begin{equation}\label{eq:indices}
    i_{\text{ue}} = D \cdot epi_{\text{idx}} + av_{\text{idx}}    
\end{equation}

\noindent
where \(epi_{\text{idx}}\) is the vector of exposure per impression indices for historical clicks, \(av_{\text{idx}}\) is the vector of avoidance indices for historical clicks, and \(D\) is a constant representing the number of distinct values. The \textit{user engagement embedding} is then obtained using an embedding layer:

\vspace{-1.5em}
\begin{equation}
\bm{ue} = \mathbf{W}_{\text{ue}}(i_{\text{ue}})
\end{equation}

\noindent where \(\mathbf{W}_{\text{ue}}\) is an embedding layer that maps each index to an embedding vector, and \(i_{\text{ue}}\) are the computed indices from the above equation \ref{eq:indices}. The \textit{user engagement embedding} (\(\bm{ue}\)) captures the spatial representation based on the quantized values of avoidance and exposure per impression. This embedding is used as input for our models to capture the information conveyed when a specific news article is largely avoided by many users but still clicked by a particular user. The concept behind the\textit{ user engagement embedding} (\(\mathbf{ue}\)) is to integrate \textbf{avoidance-awareness} into recommender models. Figure \ref{fig:grid} illustrates this division for \( D = 5 \), resulting in \( D \times D = 25 \) distinct regions.

\section{Methodology.}

Section \ref{subsec:rec_aspects} discusses various aspects of news recommendations. We introduce a new aspect: \textbf{Avoidance}. Our method, \textit{AWRS}, unlike traditional methods based only on clicks and popularity, integrates \textit{avoidance awareness}. The conceptual framework is shown in Figure \ref{fig:awrs_schema}.

\subsection{Problem Formulation.}

Given a user \( u \) and a candidate news article \( n_c \), our goal is to compute an interest score \( Int_{s} \) that quantifies user \( u \)'s potential engagement with \( n_c \). We evaluate a collection of candidate news articles \( \mathbf{N}_c = [n_{c1}, n_{c2}, \ldots, n_{cL}] \) and recommend the highest-ranking articles to user \( u \). User \( u \) has a history of clicked news articles \( \mathbf{H}_u = [h_{1}, h_{2}, \ldots, h_{M}] \). Each news article is characterized by its title (\(\mathcal{T}\)), abstract (\(\mathcal{A}\)), category (\(\mathcal{C}_{cat}\)), and associated entities (\(\mathbf{E}_i = [e_1, e_2, \ldots, e_k]\)).

\subsection{Document Encoder.}\label{ref:doc_encoder}

Our news encoder, inspired by \cite{qi2021pprec, qi2022news, wu2019neural}, employs a multi-head attention mechanism to extract information from embeddings generated by Pretrained Language Models (PLMs) \cite{iana2024train}. We use RoBERTa Base \cite{liu2019roberta} for \textit{MIND-small}, NB-BERT Base \cite{kummervold-etal-2021-operationalizing} for \textit{Adressa one-week}, and fine-tuned \textit{Japanese DeBERTa V2}\footnote{\url{https://huggingface.co/ku-nlp/deberta-v2-base-japanese}} for \textit{Nikkei one-week}, with only the last four PLM layers fine-tuned. The final news embedding $\textbf{n}$ combines title ($\textbf{n}_t$), category ($\textbf{n}_c$), and entity ($\textbf{n}_e$) embeddings, while for \textit{Nikkei one-week} and \textit{Adressa one-week}, only title and category embeddings are used.

\subsection{Avoidance-aware Relevance Predictor.}\label{sec:aw_rel}

As discussed in Section \ref{sec:relevance}, measuring relevance at a given time \( t \) involves considering several factors: the content of the article, the time elapsed since publication, avoidance and exposure per impression values, and the number of clicks. Therefore, the \textit{Avoidance-aware Relevance Predictor}, illustrated in Figure \ref{fig:model_arch} (a), evaluates the relevance of each previously clicked news article \( n \). 

This assessment incorporates the time elapsed, encoded using the Time2Vec \cite{kazemi2019time2vec} model, resulting in the time elapsed embedding \( \bm{t}_{el} \). The news embedding \( \bm{n} \), obtained from the news encoder described in Section \ref{ref:doc_encoder}, along with the number of clicks at time \( t \) denoted as \( n_{clk}(n, t) \), and the impact of avoidance and exposure per impression are integrated into the user engagement embedding \( \bm{ue} \), as detailed in Section \ref{sec:usr_eng}. Consequently, the avoidance-aware relevance scores (\( \textbf{r}_{aw} \)) for each news article are computed as follows:

\begin{figure*}[t]
  \includegraphics[width=\linewidth, height=5.5cm]{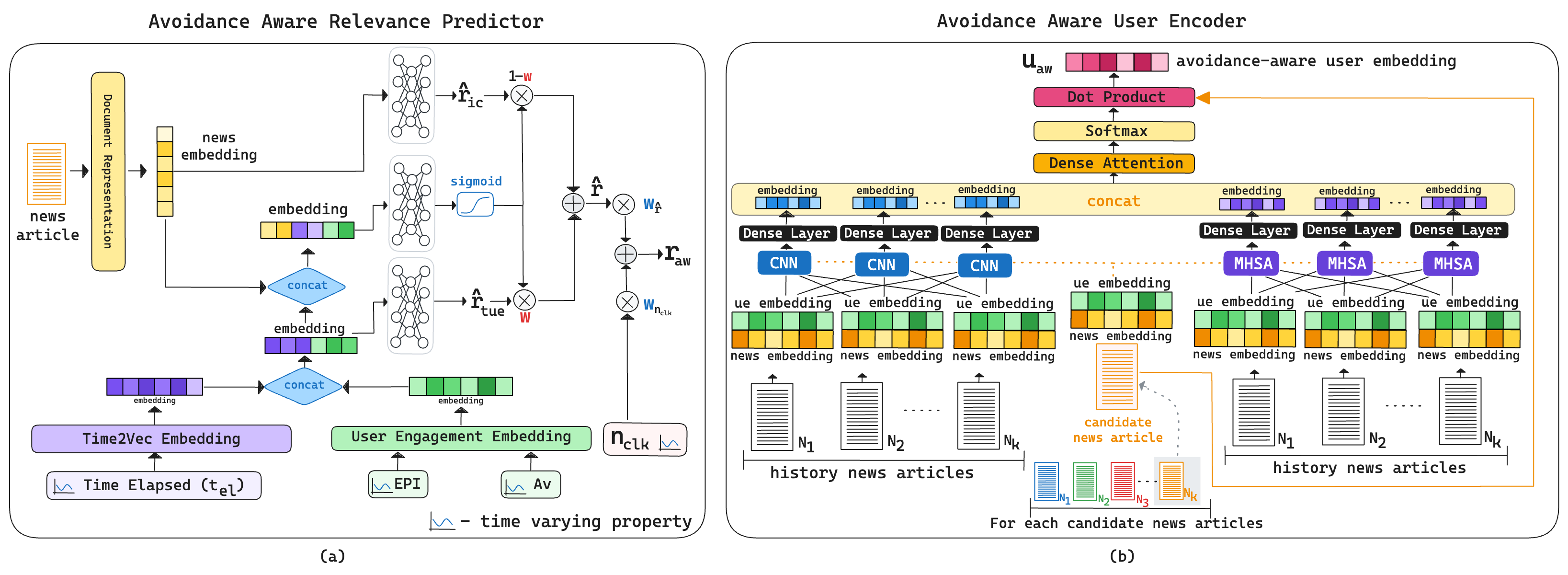}
  \caption{(a) \textit{Avoidance-aware Relevance Predictor} and (b) \textit{Avoidance-aware User Encoder} Schematics.}
  \label{fig:model_arch}
\end{figure*}

\vspace{-1.5em}
\begin{equation}\label{eq:aw_scores}
\textbf{r}_{aw} = \sigma \left( n_{\textit{clk}}(n, t) \cdot w_{\text{ctr}} + \hat{r} \cdot w_{\hat{r}} \right)
\end{equation}

\noindent
here, \(\hat{r}\) is the weighted sum of the relevance scores influenced by the news article conveyed information, defined as  $\hat{r} = W \cdot \hat{r}_{\text{ic}} + (1 - W) \cdot \hat{r}_{\text{tue}}$. Where the weight \(W\) is computed as \(W = \sigma (\Psi_1([\textbf{n}, \bm{ue}, \bm{t}_{el}]))\), where \([\cdot, \cdot, \cdot]\) denotes vector concatenation, \(\sigma\) is the sigmoid function, and \(\Psi_1\) is a dense layer. 

The term \(\hat{r}_{\text{ic}}\) represents the relevance score influenced by the news article embedding (\(\textbf{n}\)):

\vspace{-1.5em}
\begin{equation}
\hat{r}_{\text{ic}} = \Psi_2(\textbf{n}), \quad \hat{r}_{\text{tue}} = \Psi_3([\bm{ue}, \textbf{t}_{el}])
\end{equation}

\noindent
here the term \(\hat{r}_{\text{tue}}\) represents the relevance score influenced by the combination of the \textit{user engagement embedding} ($\bm{ue}$) and the time elapsed embedding ($\bm{t}_{el}$). 

Finally, the overall relevance score ($\bm{r}_{aw}$) is obtained by considering both the number of clicks up to time \(t\) (\(n_{clk}(n, t)\)) and the computed relevance score (\(\hat{r}\)), adjusted by their respective weights (\(w_{n_{clk}}\) and \(w_{\hat{r}}\)) and passed through a sigmoid activation function (\(\sigma\)) for normalization, as indicated by Equation \ref{eq:aw_scores}.

\subsection{Avoidance-aware User Encoder.}\label{sec:avoidance_aware_user_encoder}

The core concept of \textit{AWRS} is \textit{avoidance-awareness}. Building on CAUM's architecture \cite{qi2022news}, which focuses on capturing the relationship between candidate and previously clicked news, \textit{AWRS} enhances candidate-aware attention with \textit{user engagement embeddings}. This addition provides a deeper understanding of user preferences by analyzing news categories, \textit{exposure}, and \textit{avoidance}, resulting in a more refined user interest representation.

\subsubsection{Historical and Candidate News Vector Concatenation.} 

For each historical and candidate news item, we enrich its representation by appending the corresponding \textit{user engagement embedding}, resulting in enhanced feature vectors that support a more personalized recommendation approach. Given a set $\mathbf{N}_{c}$ of candidate news articles and a set $\mathbf{H}_{u}$ of historical clicks for a user $u$, we concatenate the respective \textit{user engagement embedding} $\bm{ue}$ for each item. 

Let $\mathscr{H}$ and $\mathscr{N}$ denote the sets of historical and candidate news articles with their \textit{user engagement embeddings} concatenated. Formally, for each $i$-th item:

\vspace{-0.4cm}
\begin{equation}
\bm{h}_i = [\textbf{h}_{u_i}, \bm{ue}_{h_i}]
\quad \text{and} \quad
\bm{n}_i = [\mathbf{n}_{c_i}, \bm{ue}_{c_i}]
\end{equation}

\noindent where $\bm{ue}_{h_i}$ and $\bm{ue}_{c_i}$ represent the \textit{user engagement embeddings} for the $i$-th historical and candidate news items, respectively, and the operator $[\cdot, \cdot]$ denotes vector concatenation. Thus, the resulting sets are defined as:

\vspace{-0.4cm}
\begin{equation}
\mathscr{H} = [\bm{h}_1, \dots, \bm{h}_M]
\quad \text{and} \quad
\mathscr{N} = [\bm{n}_1, \dots, \bm{n}_L]
\end{equation}

\noindent which now incorporate \textit{avoidance} information into their representations for each historical and candidate item.


\subsubsection{Avoidance-aware Self-Attention Layer.}

Given the user's click history $\mathscr{H}_{u}^{M}$, where $M$ is the total number of previously clicked articles, we employ multiple self-attention heads to capture the \textbf{relatedness} between the $i$-th and $j$-th historical clicks, $\textbf{h}_i$ and $\textbf{h}_j$, respectively. Specifically, we compute:

\vspace{-0.5cm}
\begin{equation}
    \hat{r}^k_{i, j} = \textbf{q}^T_i \textbf{W}^r_k \bm{h}_j, \quad  \textbf{q}^T_i = \textbf{Q}_u \bm{h}_i
\end{equation}

\noindent where $\hat{r}^k_{i,j}$ represents the attention score from the $k$-th head, $\textbf{Q}_u$ is the projection matrix, and $\textbf{W}^r_k$ denotes the parameters of the $k$-th attention head. This approach captures relatedness not only from the news content but also from the user engagement information.

We adaptively select significant long-range relatedness metrics to model user interest in the candidate news \(\bm{n}_c \in \mathscr{N}\) based on their contextual relevance $r^k_{i,j} = \hat{r}^k_{i,j} + \textbf{q}^T_c \textbf{W}^r_k \bm{h}_j$ where \(\textbf{q}^T_c = \textbf{Q}_c \bm{n}_c\). This enhances the attention score with candidate-specific adjustments. The augmented representation \(\textbf{l}^k_i\) for each click is formulated through attention weights:

\vspace{-1.5em}
\begin{equation}
    \textbf{l}^k_i = \textbf{W}^k_o \sum_{j=1}^N \gamma^k_j \bm{h}_j, \quad \gamma^k_j = \frac{\exp(r^k_{i,j})}{\sum_{p=1}^N \exp(r^k_{i,p})}
\end{equation}


\noindent
\(\textbf{W}^k_o\) is the projection matrix for the \(k\)-th head. The comprehensive contextual representation \(\textbf{l}_i\) for each click is derived by merging outputs from all \(K\) attention heads. 

Note that by incorporating \textit{user engagement embeddings}, our \textbf{relatedness} information now considers the influence of both \textit{avoidance} and \textit{exposure}, providing more context to user click patterns.

\subsubsection{Avoidance-aware CNN Layer.}

The historical news vectors with their respective user engagement embeddings are processed through a convolutional neural network (CNN) with self-attention mechanisms. As per \cite{qi2022news}, multiple filters capture patterns in local contexts of adjacent clicks and candidate news, represented as $\textbf{s}_i = \textbf{W}_{awc}[\bm{h}_{i-h}; \dots; \bm{h}_i; \dots; \bm{h}_{i+h}; \bm{n}_c]$. Here, \(\textbf{s}_i\) is the local contextual representation of the \(i\)-th click, \(2h + 1\) is the CNN window size, and \(\textbf{W}_{awc}\) represents the \textit{avoidance-aware} filter parameters. These representations \([s_1, s_2, ..., s_N]\) encode candidate-aware short-term user interests. A unified representation \(\textbf{m}_i\) for each click is obtained by aggregating \(\textbf{l}_i\) and \(\textbf{s}_i\) via a dense layer, \(\textbf{m}_i = \Phi_1[\textbf{s}_i, \textbf{l}_i]\).

\begin{table*}[ht]
    \centering
    \tiny
    \setlength{\tabcolsep}{2pt}
    \caption{Performance comparison for the \textit{Nikkei one-week}, \textit{MIND-small}, and \textit{Adressa one-week} datasets. The best results are \textbf{bolded}, and the second best are \underline{underlined}. All the results reported here used $D = 5$.}
    \begin{tabular}{l|c|c|c|c|c|c|c|c|c|c|c|c}
        \toprule
        & \multicolumn{4}{c|}{\textbf{\textit{Nikkei one-week}}} & \multicolumn{4}{c|}{\textbf{\textit{MIND-small}}} & \multicolumn{4}{c}{\textbf{\textit{Adressa one-week}}} \\
        \cmidrule(lr){2-5} \cmidrule(lr){6-9} \cmidrule(lr){10-13}
        \textbf{MODEL} & \textbf{AUC} & \textbf{MRR} & \textbf{nDCG@5} & \textbf{nDCG@10} & 
                         \textbf{AUC} & \textbf{MRR} & \textbf{nDCG@5} & \textbf{nDCG@10} & 
                         \textbf{AUC} & \textbf{MRR} & \textbf{nDCG@5} & \textbf{nDCG@10} \\
        \midrule
        NRMS       & 51.87$\pm$2.3 & 30.55$\pm$1.9 & 20.42$\pm$2.7 & 29.99$\pm$2.5 & 
                     49.64$\pm$0.7 & 24.37$\pm$4.4 & 22.20$\pm$4.7 & 28.80$\pm$4.8 & 
                     55.65$\pm$5.1 & 21.88$\pm$1.6 & 16.49$\pm$3.1 & 26.11$\pm$5.2 \\
        NAML       & 50.11$\pm$0.0 & 33.23$\pm$1.0 & 23.24$\pm$0.8 & 33.22$\pm$0.9 & 
                     48.54$\pm$2.5 & 26.88$\pm$4.1 & 24.84$\pm$4.4 & 31.17$\pm$4.4 & 
                     50.00$\pm$0.0 & 23.52$\pm$0.7 & 21.81$\pm$1.7 & 29.43$\pm$0.0 \\
        LSTUR      & 50.36$\pm$1.2 & 29.03$\pm$3.6 & 19.61$\pm$4.1 & 29.29$\pm$4.1 & 
                     50.01$\pm$0.0 & 29.31$\pm$1.2 & 27.50$\pm$1.4 & 33.95$\pm$1.3 & 
                     54.87$\pm$3.2 & 32.54$\pm$2.8 & 30.31$\pm$3.2 & 36.92$\pm$3.0 \\
        TANR       & 50.02$\pm$0.0 & 29.15$\pm$0.8 & 20.39$\pm$0.2 & 29.81$\pm$0.3 & 
                     52.24$\pm$1.5 & 29.41$\pm$1.9 & 27.73$\pm$1.5 & 34.25$\pm$1.4 & 
                     52.39$\pm$4.1 & 24.25$\pm$4.4 & 22.65$\pm$5.0 & 29.67$\pm$4.2 \\
        SentiRec   & 50.00$\pm$0.0 & 29.46$\pm$0.9 & 20.51$\pm$0.1 & 29.96$\pm$0.0 & 
                     52.89$\pm$2.9 & 27.73$\pm$1.3 & 26.42$\pm$1.3 & 32.81$\pm$1.3 & 
                     52.44$\pm$1.4 & 20.24$\pm$1.2 & 16.80$\pm$1.7 & 26.85$\pm$1.5 \\
        MINER      & \underline{55.71$\pm$1.1} & 31.73$\pm$1.1 & 23.04$\pm$0.3 & \underline{34.65$\pm$1.3} & 
                     61.41$\pm$0.7 & 27.26$\pm$0.5 & 25.91$\pm$0.7 & 32.74$\pm$0.5 & 
                     54.43$\pm$5.2 & 26.40$\pm$7.2 & 22.55$\pm$11.2 & 31.41$\pm$11.0 \\
        MINS       & 50.68$\pm$1.3 & 27.78$\pm$1.0 & 17.21$\pm$0.8 & 26.58$\pm$0.3 & 
                     47.60$\pm$4.3 & 24.72$\pm$4.3 & 23.00$\pm$4.8 & 29.28$\pm$4.8 & 
                     69.88$\pm$0.5 & 35.34$\pm$0.4 & 36.35$\pm$3.5 & 44.35$\pm$4.6 \\
        CenNewsRec & 52.68$\pm$0.2 & 31.20$\pm$2.1 & 20.97$\pm$3.6 & 29.22$\pm$3.5 & 
                     49.46$\pm$0.9 & 20.88$\pm$0.8 & 18.59$\pm$0.8 & 24.92$\pm$0.9 & 
                     53.99$\pm$5.1 & 28.86$\pm$2.9 & 31.46$\pm$3.9 & 36.32$\pm$1.5 \\
        MANNeR-CR  & 50.00$\pm$0.0 & \underline{33.74$\pm$0.4} & \underline{24.53$\pm$0.3} & 34.93$\pm$0.4 & 
                     \underline{62.23$\pm$2.4} & 30.72$\pm$1.8 & 28.81$\pm$1.8 & 35.43$\pm$1.7 & 
                     53.63$\pm$6.3 & 25.66$\pm$3.5 & 22.40$\pm$5.0 & 30.62$\pm$5.1 \\
        PP-REC     & 49.70$\pm$1.5 & 27.31$\pm$1.2 & 18.70$\pm$0.1 & 28.09$\pm$0.2 & 
                     56.86$\pm$1.6 & 29.12$\pm$1.0 & 26.95$\pm$1.1 & 33.01$\pm$1.0 & 
                     \underline{72.83$\pm$5.1} & \underline{52.53$\pm$9.1} & \underline{51.25$\pm$9.9} & \underline{55.94$\pm$8.4} \\
        CAUM       & 55.27$\pm$2.2 & 32.89$\pm$2.2 & 23.51$\pm$2.8 & 32.76$\pm$3.5 & 
                     55.76$\pm$0.7 & \underline{31.68$\pm$0.6} & \underline{30.04$\pm$0.6} & \underline{36.40$\pm$0.6} & 
                     60.87$\pm$4.4 & 25.66$\pm$1.8 & 25.20$\pm$2.2 & 31.74$\pm$6.0 \\
        \textbf{\textit{AWRS}} & \textbf{65.75$\pm$1.3} & \textbf{48.95$\pm$1.4} & \textbf{39.07$\pm$1.0} & \textbf{46.81$\pm$0.9} & 
                                  \textbf{63.70$\pm$1.1} & \textbf{33.39$\pm$0.9} & \textbf{31.43$\pm$0.9} & \textbf{37.84$\pm$0.8} & 
                                  \textbf{81.20$\pm$7.8} & \textbf{55.72$\pm$12.9} & \textbf{59.26$\pm$13.7} & \textbf{63.12$\pm$11.8} \\
        \bottomrule
    \end{tabular}
    \label{tab:results}
\end{table*}

\subsubsection{Avoidance-aware Final Attention Layer.}

We use a candidate-aware attention network to model the importance of clicked news based on their relevance to \textit{avoidance-aware} candidate news \( \bm{n}_c \in \mathscr{N}\). This builds the \textit{avoidance-aware} user embedding representation \(\textbf{u}_{aw}\), given by $\textbf{u}_{aw} = \sum^N_{i=1} \alpha_i \textbf{m}_i$ where \(\alpha_i\) is the weight of the \(i\)-th click:

\vspace{-1.5em}
\begin{equation}
    \alpha_i = \frac{\exp(\Phi_2(\textbf{m}_i, \bm{n}_c))}{\sum^N_{j=1} \exp(\Phi_2(\textbf{m}_j, \bm{n}_c))}
\end{equation}

\noindent
here, \(\Phi_2\) is a dense layer. This approach encodes \textit{avoidance-aware} user interests relevant to the candidate news into \(\textbf{u}_{aw}\), enhancing interest matching accuracy. The general overview of the \textit{Avoidance-aware User Encoder} is shown in Figure \ref{fig:model_arch} (b).

\subsubsection{Final Relevance Scores.}

The relevance scores combine the news candidate's \textit{avoidance-awareness} $\bm{n}_c \in \mathscr{N}$ with the score from the \textit{avoidance-aware} user vector $\textbf{u}_{aw}$. Specifically, the preliminary interest scores are calculated as $Int_{s}' = \bm{n}_c^{T} \cdot \textbf{u}_{aw}$. However, to obtain the final scores, we apply our \textit{Avoidance-aware Relevance Predictor}, which considers the time-varying nature of avoidance, as shown in Figure \ref{fig:awrs_schema}. Thus,

\vspace{-1.5em}
\begin{equation}
    Int_{s} = (1 - \eta) \cdot \mathbf{r}_{aw} + \eta \cdot Int_{s}'
\end{equation}

\noindent
where $\eta = \sigma(\Phi_3(\mathbf{u}_{aw}))$. Here, \(\Phi_3\) is a dense layer, \(\mathbf{r}_{aw}\) is the relevance \textit{avoidance-aware} score as defined earlier by equation \ref{eq:aw_scores}, and \(\eta\) is computed from the user representation \(\mathbf{u}_{aw}\) using a dense network with sigmoid activation. This approach allows our model to better capture patterns related to the information conveyed by the \textit{avoidance} and \textit{exposure} values of a news article.

\subsection{Loss Function.}

Following \cite{wu-etal-2019-neural-news}, we employ negative sampling during training. For each clicked news (positive sample), \( K \) non-clicked items (negative samples) are randomly drawn from the same impression and shuffled. The click probability of the positive item is \( \hat{y}^+ \), and the scores for the \( K \) negatives are \([\hat{y}^-_1, \hat{y}^-_2, \ldots, \hat{y}^-_K]\). These scores are normalized using:

\vspace{-1.5em}
\begin{equation}
    p_i = \frac{\exp(\hat{y}^+_i)}{\exp(\hat{y}^+_i) + \sum_{j=1}^{K} \exp(\hat{y}^-_{i,j})}    
\end{equation}

This forms a pseudo \((K+1)\)-way classification problem, where the loss function is the negative log-likelihood: $\mathcal{L} = - \sum_{i \in S} \log(p_i)$.

\section{Experiment.}

\subsection{Datasets.}\label{sec:datasets}

We evaluated \textit{AWRS} on three datasets: \textit{MIND-small} (English) \cite{wu-etal-2020-mind}, \textit{Adressa one-week} (Norwegian) \cite{10.1145/3106426.3109436}, and a proprietary Japanese dataset from Nikkei News\footnote{\url{https://www.nikkei.com/}}, based in Japan, is a leading media corporation renowned for its comprehensive coverage of business, economic, and financial news. For \textit{Nikkei one-week}, we analyzed 15,803 news articles, split into training (Jan 16-20, 2023), validation (Jan 21, 2023), and testing (Jan 22, 2023). The \textit{Adressa one-week} dataset includes 22,136 articles, with training (Jan 1-5, 2017), validation (Jan 6, 2017), and testing (Jan 7, 2017). The \textit{MIND-small} dataset contains 93,698 articles, divided into training (Nov 9-13, 2019), validation (Nov 14, 2019), and testing (Nov 15, 2019). Additionally, Appendix \ref{sec:appendix_dataset} has details on the dataset splits.

\subsection{Baselines and Evaluation Metrics.}

We conducted a comparative analysis of the \textit{AWRS} model against several state-of-the-art baseline models to evaluate its performance. The models compared include: (1) NRMS \cite{wu-etal-2019-neural-news}, (2) NAML \cite{wu2019neural}, (3) LSTUR \cite{an-etal-2019-neural}, (4) TANR \cite{wu-etal-2019-neural-news-recommendation}, (5) SentiRec \cite{wu-etal-2020-sentirec}, (6) MINER \cite{ li-etal-2022-miner}, (7) MINS \cite{wang2022modelingmultiinterestnewssequence}, (8) CenNewsRec \cite{qi-etal-2020-privacy}, (9) MANNeR-CR\footnote{For the \textit{Nikkei one-week} dataset with the MANNeR-CR model, we reduced the token limit due to memory constraints.} \cite{iana2024train}, (10) PP-REC \cite{qi2021pprec}, and (11) CAUM \cite{qi2022news}. In accordance with prior research \cite{qi2022news}, we evaluate model performance using AUC, MRR, nDCG@5 and nDCG@10.


\begin{table}[htbp]
    \centering
    \small  
    \setlength{\tabcolsep}{2pt}  
    \caption{Performance comparison on the \textit{MIND-small} and \textit{Adressa one-week} datasets using $D = 5$ and word embeddings. The best results are \textbf{bolded}, and the second best are \underline{underlined}.}
    \resizebox{\columnwidth}{!}{ 
    \begin{tabular}{l|c|c|c|c|c}
        \hline
        \textbf{Dataset} & \textbf{Model} & \textbf{AUC} & \textbf{MRR} & \textbf{nDCG@5} & \textbf{nDCG@10} \\
        \hline
        \multirow{9}{*}{\textit{MIND-small}} 
        & NRMS & 55.71$\pm$2.37 & 29.24$\pm$2.32 & 27.30$\pm$2.21 & 33.70$\pm$2.02 \\
        & NAML & 50.12$\pm$0.09 & \underline{33.88$\pm$1.94} & 32.17$\pm$1.70 & 38.43$\pm$1.69 \\
        & LSTUR & 54.52$\pm$0.58 & 32.09$\pm$0.28 & 30.14$\pm$0.36 & 36.60$\pm$0.29 \\
        & TANR & 57.66$\pm$3.26 & 32.57$\pm$0.15 & 30.68$\pm$0.25 & 36.95$\pm$0.11 \\
        & SentiRec & 54.03$\pm$0.47 & 30.51$\pm$0.07 & 28.23$\pm$0.02 & 34.78$\pm$0.06 \\
        & CenNewsRec & 53.40$\pm$1.44 & 26.01$\pm$0.32 & 24.61$\pm$0.25 & 31.32$\pm$0.19 \\
        & CAUM & \underline{60.91$\pm$0.99} & 33.77$\pm$0.47 & 31.69$\pm$0.47 & 38.19$\pm$0.44 \\
        & GLORY & 52.15$\pm$0.13 & 29.88$\pm$0.22 & \underline{32.25$\pm$0.17} & \underline{38.48$\pm$0.08} \\
        & \textbf{\textit{AWRS}} & \textbf{66.21$\pm$0.56} & \textbf{34.15$\pm$0.27} & \textbf{32.51$\pm$0.17} & \textbf{38.96$\pm$0.23} \\
        \hline
        \multirow{9}{*}{\textit{Adressa one-week}} 
        & NRMS & 59.55$\pm$2.85 & 22.76$\pm$2.63 & 20.50$\pm$4.30 & 30.59$\pm$2.81 \\
        & NAML & 50.18$\pm$0.32 & 23.54$\pm$3.15 & 21.36$\pm$4.99 & 31.38$\pm$5.89 \\
        & LSTUR & 64.58$\pm$1.04 & \underline{27.56$\pm$0.95} & \underline{28.02$\pm$1.50} & \underline{35.83$\pm$0.18} \\
        & TANR & 55.33$\pm$1.27 & 21.19$\pm$1.00 & 17.30$\pm$1.78 & 27.36$\pm$1.08 \\
        & SentiRec & 47.85$\pm$0.40 & 19.43$\pm$0.14 & 15.83$\pm$0.47 & 21.73$\pm$0.00 \\
        & CenNewsRec & 58.07$\pm$5.57 & 21.36$\pm$5.57 & 19.06$\pm$8.13 & 27.56$\pm$7.23 \\
        & CAUM & \underline{64.73$\pm$3.86} & 24.85$\pm$3.17 & 23.22$\pm$4.89 & 33.21$\pm$4.57 \\
        & GLORY & 57.52$\pm$5.59 & 20.24$\pm$0.80 & 18.52$\pm$1.11 & 27.61$\pm$2.66 \\
        & \textbf{\textit{AWRS}} & \textbf{67.71$\pm$7.88} & \textbf{34.74$\pm$4.18} & \textbf{35.55$\pm$5.17} & \textbf{41.62$\pm$6.80} \\
        \hline
    \end{tabular}}
    \label{tab:performance_metrics}
\end{table}

\vspace{-0.5cm}
\subsection{Environment Configuration.}

During training, we utilized mixed precision with the Adam optimizer \cite{kingma2017adam}. The learning rates were set to 1e-5 for \textit{MIND-small} and \textit{Nikkei one-week}, and 1e-6 for \textit{Adressa one-week}, with \( k = 4 \) negative samples. Models were trained for a maximum of 10 epochs for \textit{MIND-small} with early stopping configured to halt training after 3 epochs without improvement. For \textit{Nikkei one-week} and \textit{Adressa one-week} the models were trained A single NVIDIA A100 GPU was used. Our model is implemented based on the newsreclib~\footnote{\url{https://github.com/andreeaiana/newsreclib}} and is available as the awrs\_recsys~\footnote{\url{https://github.com/toyolabo/awrs_recsys}}. Avoidance and user engagement embeddings were computed hourly for \textit{MIND-small}, every two hours for \textit{Nikkei one-week}, and every five hours for \textit{Adressa one-week}. The reported results were obtained by running each model three times and calculating the average and standard deviation.

\section{Results.}

We proceed with a comprehensive evaluation of \textit{AWRS} by addressing the following research questions (RQ).

\subsection{\textbf{RQ1} (Accuracy).} \textit{Does the proposed \textit{AWRS} method outperform existing baselines?} 

Table \ref{tab:results} compares the performance of our proposed \textit{AWRS} model with baseline models. As shown, \textit{AWRS} achieves notable improvements across most key metrics, highlighting its enhanced awareness of avoidance and exposure patterns. For instance, a user's choice to read an article about a niche topic like chess—often widely avoided—suggests a strong personal interest. In contrast, an article on the Oscars, which attracts broad attention, may be clicked due to its popularity rather than genuine interest. These patterns evolve over time; during a period of controversy, chess articles may gain traction, whereas interest in the Oscars may decline when far from the event’s date. This underscores the strong connection between \textbf{avoidance} and \textbf{time}.

The baseline models employ sophisticated architectures to capture user behavior effectively. However, by integrating \textit{user engagement embeddings} with our \textit{Avoidance-aware User Encoder} and \textit{Avoidance-aware Relevance Predictor}, as illustrated in Figures \ref{fig:awrs_schema} and \ref{fig:model_arch}, \textit{AWRS} provides a deeper understanding of user preferences. This enables a more \textit{avoidance-aware} contextual assessment of each article, capturing both user interests and avoidance behavior more effectively.

\subsubsection{\textbf{RQ2} (Performance without PLMs).} \textit{How does our model perform in the absence of pretrained language models (PLMs)? }

We investigated how excluding pretrained language models (PLMs) in favor of word embeddings GloVe embeddings \cite{pennington-etal-2014-glove} for Egnlish and BPEmb \cite{heinzerling2018bpemb} for Norwegian would affect our model's performance. For evaluation, we utilized the same datasplit as described in Section \ref{sec:datasets}. In this comparison, we included several baseline models for a more comprehensive evaluation: (1) NRMS \cite{wu-etal-2019-neural-news}, (2) NAML \cite{wu2019neural}, (3) LSTUR \cite{an-etal-2019-neural}, (4) TANR \cite{wu-etal-2019-neural-news-recommendation}, (5) SentiRec \cite{wu-etal-2020-sentirec}, (6) CenNewsRec \cite{qi-etal-2020-privacy}, (7) CAUM \cite{qi2022news}, and (8) GLORY \cite{Yang_2023}. Table \ref{tab:performance_metrics} demonstrates that, overall, \textit{AWRS} exhibits superior performance across all key metrics. It is important to address the slight performance differences between GLORY \cite{Yang_2023} and other models. These differences may arise from the variation in time splits used. We trained all models for 20 epochs with a early stopping configured to stop training after 5 epochs without improvement.

\subsubsection{\textbf{RQ3} (Effect of Each Component).} \textit{What is the impact of each AWRS component on accuracy scores?} 

To evaluate the influence of each component in our architecture, specifically the \textit{Avoidance-aware User Encoder} and the \textit{Avoidance-aware Relevance Predictor}, we conducted experiments by removing each component separately for all datasets. As shown in Table \ref{tab:awrs_scores}, each component performs worse individually than when combined. This occurs because the \textit{Avoidance-aware User Encoder} captures historical patterns of user behavior regarding avoidance, such as a preference for more or less avoided news articles. Meanwhile, the \textit{Avoidance-aware Relevance Predictor} understands the temporal and global impact of avoidance, adjusting relevance based on the timing of recommendations. This dual approach improves the model's accuracy in predicting user preferences.

\begin{table}[htbp]
    \centering
    \small  
    \setlength{\tabcolsep}{2pt}  
    \caption{Effect of the \textit{Avoidance-aware User Encoder}, denoted as ``rel," and the \textit{Avoidance-aware Relevance Predictor}, denoted as ``avoid," components on \textit{AWRS}.}
    \resizebox{\columnwidth}{!}{ 
    \begin{tabular}{l|c|c|c|c|c}
        \hline
        \textbf{Dataset} & \textbf{\textit{AWRS} Model} & \textbf{AUC} & \textbf{MRR} & \textbf{nDCG@5} & \textbf{nDCG@10} \\
        \hline
        \multirow{3}{*}{\textit{MIND-small}} 
        & only rel & 58.18 & 32.44 & 30.83 & 36.90 \\
        & only avoid & 58.38 & 30.77 & 28.84 & 35.11 \\
        & \textbf{avoid + rel} & \textbf{63.70} & \textbf{33.39} & \textbf{31.43} & \textbf{37.84} \\
        \hline
        \multirow{3}{*}{\textit{Adressa one-week}} 
        & only rel & 63.54 & 36.21 & 39.48 & 45.26 \\
        & only avoid & 54.23 & 25.64 & 22.04 & 24.99 \\
        & \textbf{avoid + rel} & \textbf{81.20} & \textbf{55.72} & \textbf{59.26} & \textbf{63.12} \\
        \hline
        \multirow{3}{*}{\textit{Nikkei one-week}} 
        & only rel & 63.62 & 46.38 & 37.01 & 45.34 \\
        & only avoid & 65.22 & 48.89 & 37.84 & 45.68 \\
        & \textbf{avoid + rel} & \textbf{65.75} & \textbf{48.95} & \textbf{39.06} & \textbf{46.81} \\
        \hline
    \end{tabular}}
    \label{tab:awrs_scores}
\end{table}

\vspace{-0.5cm}
\subsubsection{\textbf{RQ4} (Effect of Varying $D$).} \textit{How does accuracy change with different values of \(D\)?}

To assess the impact of the parameter \( D \) on the \textit{AWRS} model, we conducted experiments with values \( D = 5, 7, 10, 15, \) and \( 20 \). As shown in Table \ref{tab:varying_D}, adjusting \( D \) significantly influences model performance, albeit with nuanced effects. For the \textit{Nikkei one-week} dataset, the best results are observed at \( D = 5 \) and \( D = 7 \), while the \textit{Adressa one-week} dataset achieves optimal performance at \( D = 5 \). In the case of the \textit{MIND-small} dataset, the highest AUC score is attained at \( D = 20 \), MRR peaks at \( D = 10 \), and both nDCG@5 and nDCG@10 reach their highest values at \( D = 15 \). These findings highlight the importance of conducting a grid search to identify the most effective \( D \) values for each dataset.

\begin{table}[htbp]
    \centering
    \small  
    \setlength{\tabcolsep}{2pt}  
    \caption{\textit{AWRS} performance for different $D$ values.}
    \resizebox{\columnwidth}{!}{ 
    \begin{tabular}{l|c|c|c|c|c|c}
        \hline
        \textbf{Dataset} & \textbf{Metrics} & \textbf{D = 5} & \textbf{D = 7} & \textbf{D = 10} & \textbf{D = 15} & \textbf{D = 20} \\
        \hline
        \multirow{4}{*}{\textit{MIND-small}} 
        & AUC & 63.70 & 62.13 & 64.30 & \underline{64.79} & \textbf{64.94} \\
        & MRR & 33.39 & 32.14 & \textbf{34.99} & \underline{34.81} & 34.09 \\
        & nDCG@5 & 31.43 & 30.49 & \underline{32.85} & \textbf{32.89} & 32.26 \\
        & nDCG@10 & 37.84 & 36.84 & \underline{39.14} & \textbf{39.23} & 38.64 \\
        \hline
        \multirow{4}{*}{\textit{Adressa one-week}} 
        & AUC & \textbf{81.20} & 65.61 & 69.97 & 64.43 & \underline{76.52} \\
        & MRR & \textbf{55.72} & 43.10 & 45.98 & 41.09 & \underline{50.55} \\
        & nDCG@5 & \textbf{59.26} & 41.72 & 47.99 & 38.69 & \underline{53.00} \\
        & nDCG@10 & \textbf{63.12} & 46.70 & 50.62 & 44.59 & \underline{60.33} \\
        \hline
        \multirow{4}{*}{\textit{Nikkei one-week}} 
        & AUC & \textbf{65.75} & \underline{64.66} & 62.52 & 62..83 & 64.24 \\
        & MRR & \textbf{48.95} & \underline{48.04} & 41.68 & 44.97 & 44.94 \\
        & nDCG@5 & \textbf{39.06} & \underline{37.32} & 31.78 & 34.05 & 34.35 \\
        & nDCG@10 & \textbf{46.81} & \underline{45.31} & 40.96 & 42.79 & 43.30 \\
        \hline
    \end{tabular}}
    \label{tab:varying_D}
\end{table}



\vspace{-1cm}
\section{Conclusion.}

\subsection{Architecture Choices}

As discussed in Section \ref{sec:avoidance_aware_user_encoder}, CAUM \cite{qi2022news} was selected for its ability to capture relationships between candidate and previously viewed news. By incorporating \textit{user engagement embeddings} through the \textit{Avoidance-aware User Encoder} and the \textit{Avoidance-aware Relevance Predictor}, \textit{AWRS} enhances the original model by aligning content similarity, user preferences, and avoidance patterns. We believe that the \textit{user engagement embedding} $\bm{ue}$ can be implemented on top of other models, presenting a promising avenue for future research. Our experiments indicate that CAUM and \textit{AWRS} exhibit comparable training, validation, and testing times, with the primary drawback of the latter being increased dataset loading times due to the additional avoidance-related information.

\subsection{Summary.}

This study underscores the importance of \textit{avoidance awareness} in enhancing the predictive accuracy of recommender systems. By analyzing timely avoidance behaviors, we gain deeper insights into user preferences, allowing us to incorporate this context into our models. Our \textit{AWRS} model, integrating the \textit{Avoidance-aware Relevance Predictor} and \textit{Avoidance-aware User Encoder} modules, demonstrates strong performance across three real-world datasets, improving various benchmark metrics. This approach not only boosts accuracy but also lays the groundwork for future research on user behavior through \textbf{avoidance}.



\section*{Acknowledgments}
This work is partially supported by ``Joint Usage/Research Center for Interdisciplinary Large-scale Information Infrastructures (JHPCN)" in Japan (Project ID: jh241004), JSPS KAKENHI Grant Number 23K28098, and the Monbukagakusho: MEXT (Ministry of Education, Culture, Sports, Science and Technology - Japan) scholarship.

\appendix

\section{Dataset Statistics} \label{sec:appendix_dataset}

Table \ref{tab:dataset_stats} presents detailed statistics on the number of ``Impressions" (Denoted as Impr.) and ``Users" for each dataset used in this study: \textit{MIND-small} \cite{wu-etal-2020-mind}, \textit{Adressa one-week} \cite{10.1145/3106426.3109436}, and the proprietary dataset \textit{Nikkei one-week}. Additionally, Table \ref{tab:dataset_stats_dates} summarizes the training, validation, and testing periods for each dataset.

\begin{table}[H]
\centering
\scriptsize
\setlength{\tabcolsep}{2pt}  
\caption{Dataset Statistics}
\resizebox{\columnwidth}{!}{  
\begin{tabular}{@{}cccc@{}}
\toprule
\textbf{Dataset} & \textbf{Train} & \textbf{Validation} & \textbf{Test} \\ \midrule
 & Impr./Users & Impr./Users & Impr./Users \\ \midrule
\textit{Adressa} & 181,279 / 83,599 & 36,412 / 27,943 & 145,626 / 68,565 \\
\textit{MIND-small} & 124,229 / 45,214 & 29,498 / 19,703 & 70,938 / 48,593 \\
\textit{Nikkei one-week} & 137,142 / 23,139 & 10,560 / 6,201 & 9,695 / 5,805 \\ \bottomrule
\end{tabular}
}
\label{tab:dataset_stats}
\end{table}

\begin{table}[ht]
\centering
\scriptsize
\setlength{\tabcolsep}{2pt}  
\caption{Training, validation, and testing periods for each dataset.}
\resizebox{\columnwidth}{!}{  
\begin{tabular}{@{}lccc@{}}
\toprule
\textbf{Dataset} & \textbf{Training Period} & \textbf{Validation Period} & \textbf{Testing Period} \\ \midrule
\textit{Nikkei one-week}    & Jan 16-20, 2023   & Jan 21, 2023  & Jan 22, 2023 \\ 
\textit{Adressa one-week}   & Jan 1-5, 2017     & Jan 6, 2017   & Jan 7, 2017  \\ 
\textit{MIND-small}    & Nov 9-13, 2019    & Nov 14, 2019  & Nov 15, 2019 \\ \bottomrule
\end{tabular}
}
\label{tab:dataset_stats_dates}
\end{table}

\section{Hourly Avoidance}
\label{sec:appendix_hourly}

\begin{figure}[H]
  \centering
  \includegraphics[width=0.9\linewidth]{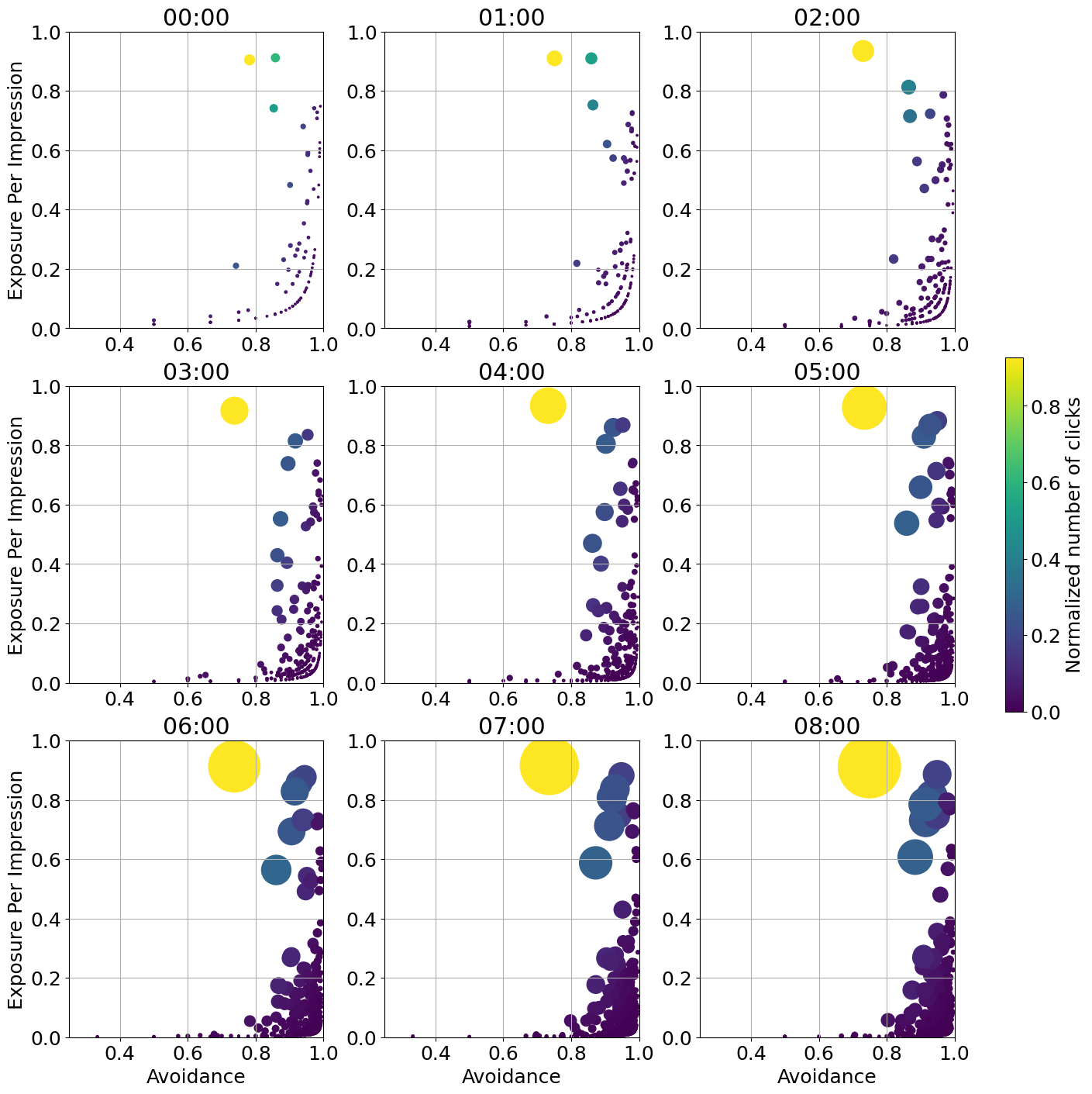}
  \caption{\(Av(n, t) \text{ vs. } EPI(n, t)\) from 00:00 to 08:00 on 11/09/2019}
  \label{fig:avoidance_00_to_08}
\end{figure}

\begin{figure}[H]
  \centering
  \includegraphics[width=0.9\linewidth]{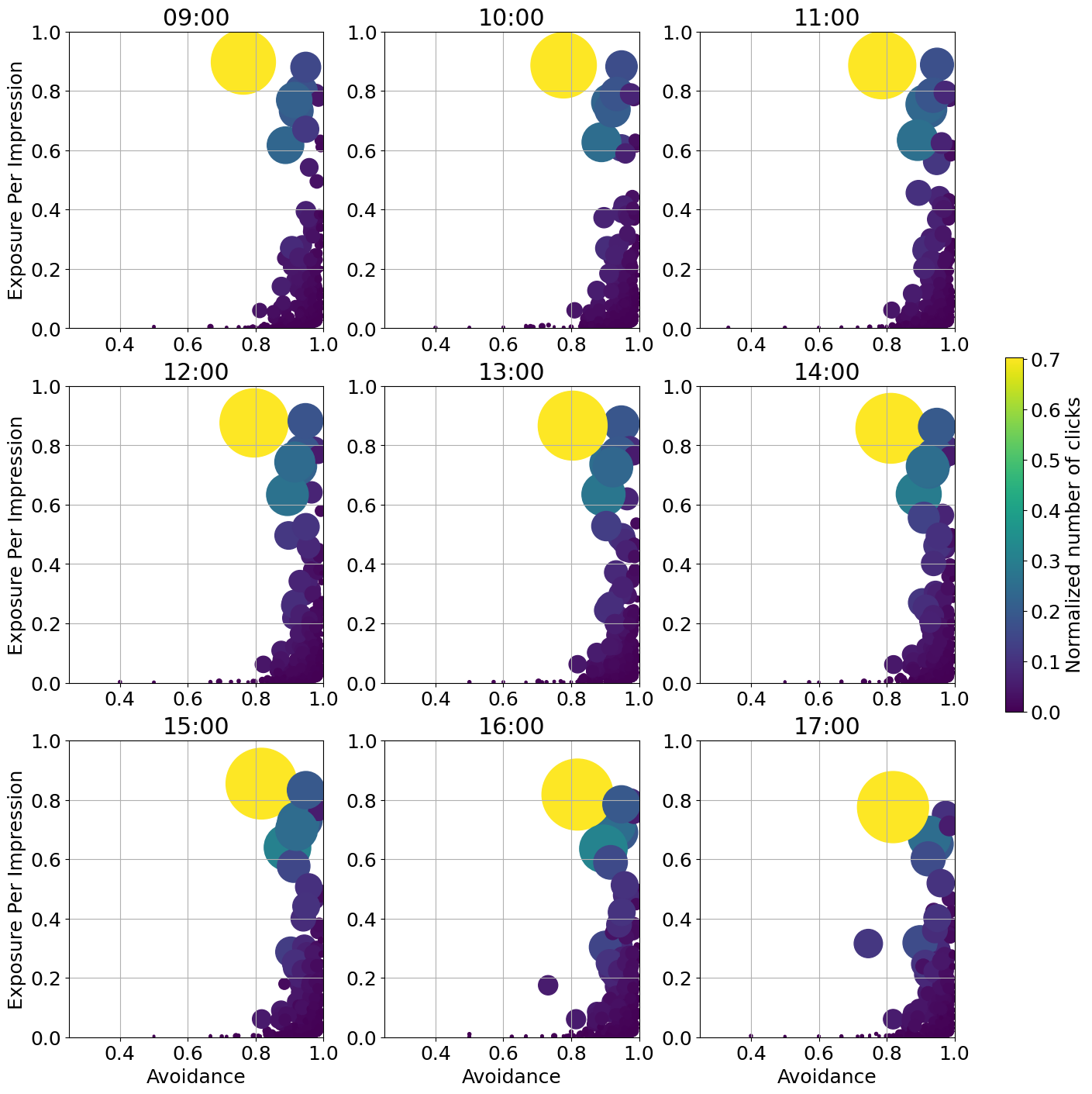}
  \caption{\(Av(n, t) \text{ vs. } EPI(n, t)\) from 09:00 to 16:00 on 11/09/2019}
  \label{fig:avoidance_08_to_16}
\end{figure}

\begin{figure}[H]
  \centering
  \includegraphics[width=0.9\linewidth]{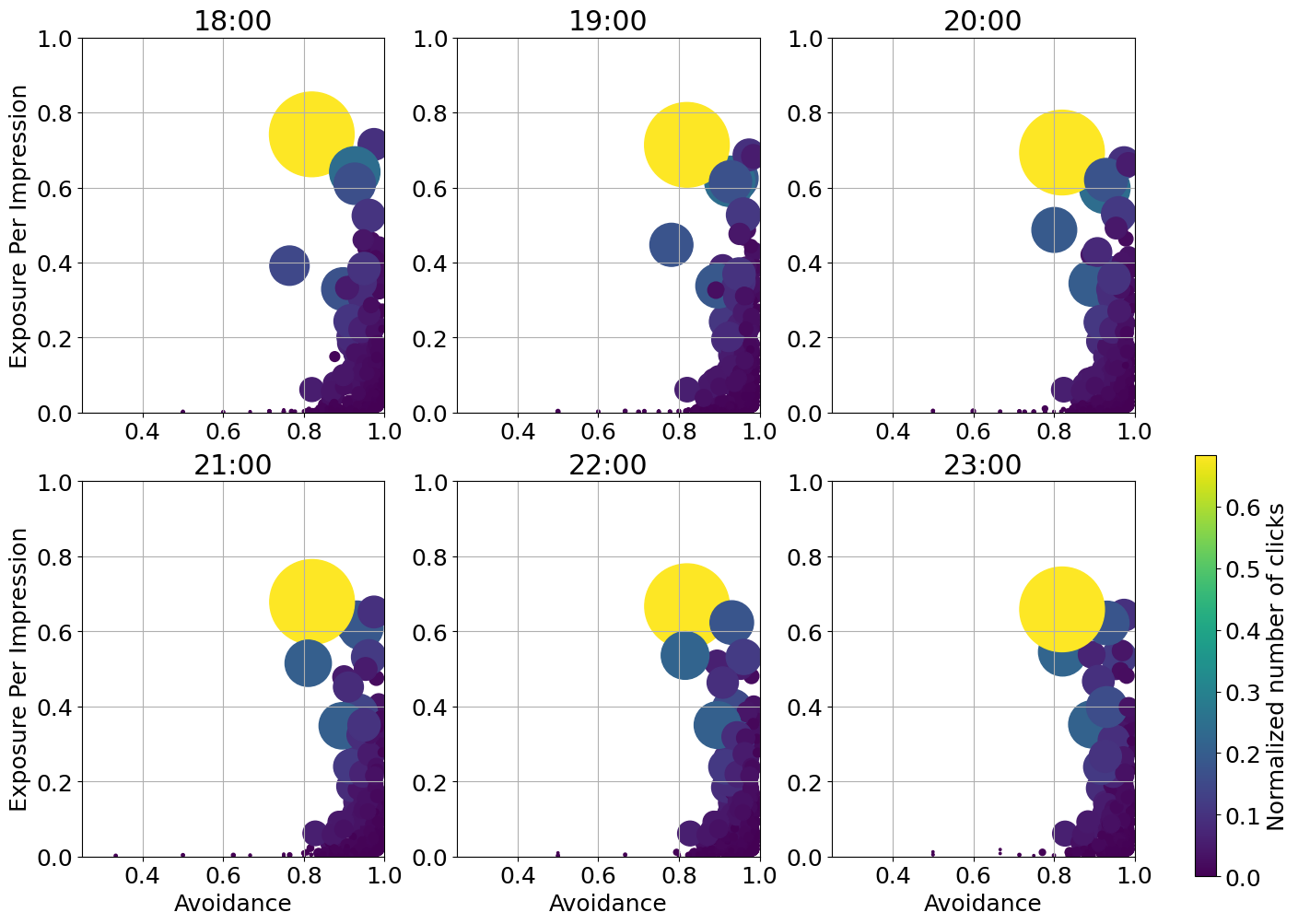}
  \caption{\(Av(n, t) \text{ vs. } EPI(n, t)\) from 17:00 to 23:00 on 11/09/2019}
  \label{fig:avoidance_16_to_23}
\end{figure}

To provide a more detailed view of the hourly change in \textit{avoidance}, the figures \ref{fig:avoidance_00_to_08}, \ref{fig:avoidance_08_to_16}, and \ref{fig:avoidance_16_to_23} present the hourly change of \( Av(n, t) \) vs \( EPI(n, t) \) for the date 11/09/2019 from 00:00 to 23:00 for the \textit{MIND-small} dataset.

\end{document}